\documentclass[fleqn,10pt]{SelfArx} 

\usepackage{caption}
\usepackage{subcaption}

\definecolor{color1}{RGB}{0,0,90} 
\definecolor{color2}{RGB}{0,20,20} 

\usepackage{hyperref} 
\hypersetup{hidelinks,colorlinks,breaklinks=true,urlcolor=color2,citecolor=color1,linkcolor=color1,bookmarksopen=false,pdftitle={Title},pdfauthor={Author}}

\JournalInfo{arXiv} 
\Archive{} 

\PaperTitle{Sub-micron period lattice structures of magnetic microtraps for ultracold atoms on an atom chip} 

\Authors{I.\ Herrera$^{1}$*, Y.\ Wang$^{1}$, P.\ Michaux$^{2}$, D. Nissen$^{4}$, P.\ Surendran$^{1}$, S. Juodkazis$^{2}$, S.\ Whitlock$^{3}$, R. J.\ McLean$^{1}$, A.\
Sidorov$^{1}$, M.\ Albrecht$^{4}$, P.\ Hannaford$^{1}$} 
\affiliation{$^{1}$\textit{Centre for Quantum and Optical Science, Swinburne University of Technology, Melbourne, Australia 3122}} 
\affiliation{$^{2}$\textit{Centre for Micro-Photonics, Swinburne University of Technology, Melbourne, Australia 3122}} 
\affiliation{$^{3}$\textit{Physikalisches Institut, Universit\"{a}t Heidelberg, Im Neuenheimer Feld 226, 69120 Heidelberg, Germany}} 
\affiliation{$^{4}$\textit{Experimental Physics IV, Institut of Physics, Universit\"{a}t Augsburg, Universit\"{a}tsstra{\ss}e 1,86135 Augsburg, Germany}} 
\affiliation{*\textbf{Ivan Herrera}: iherrerabenzaquen@swin.edu.au} 

\newcommand{\keywordname}{Keywords} 

\Abstract{We report on the design, fabrication and characterization of magnetic nanostructures to create a lattice of magnetic traps with sub--micron period for trapping ultracold atoms. These magnetic nanostructures were fabricated by patterning a Co/Pd multilayered magnetic film grown on a silicon substrate using high precision e-beam lithography and reactive ion etching. The Co/Pd film was chosen for its small grain size and high remanent magnetization and coercivity. The fabricated structures are designed to magnetically trap $^{87}$Rb atoms above the surface of the magnetic film with 1D and 2D (triangular and square) lattice geometries and sub-micron period. Such magnetic lattices can be used for quantum tunneling and quantum simulation experiments, including using geometries and periods that may be inaccessible with optical lattices.~}

\begin{document}

\flushbottom 

\maketitle 


\thispagestyle{empty} 


\section{Introduction}

The trapping of ultracold atoms in periodic lattices has gained great attention in atomic physics research  over the past decade. It has enhanced our understanding in atomic and molecular physics, as well as of condensed matter phenomena such as quantum phase transitions. 
Nearly all of the experiments in this field have been based on optical lattices~\cite{Jaksch98PRL,Morsch06,Bloch08}, in which the lattice potential is produced by the interference of intersecting laser beams. Such experiments have provided, for example, unprecedented access to studies of low-dimensional quantum gases~\cite{Paredes04,Kinoshita06}, the Josephson effect~\cite{Cataliotti01,Albiez05}, the Mott insulator to superfluid quantum phase transition~\cite{Greiner02} and applications in quantum information processing~\cite{Calarco00,Monroe07}. Particularly exciting applications arise in the context of quantum simulation~\cite{Lewenstein07}. However, optical lattices have certain limitations, such as a low degree of design flexibility, difficulty in generating arbitrary trap geometries and restrictions on the lattice spacing imposed by the optical wavelength.

\begin{figure}[h!]
 \centering
 \includegraphics[width=0.48\textwidth]{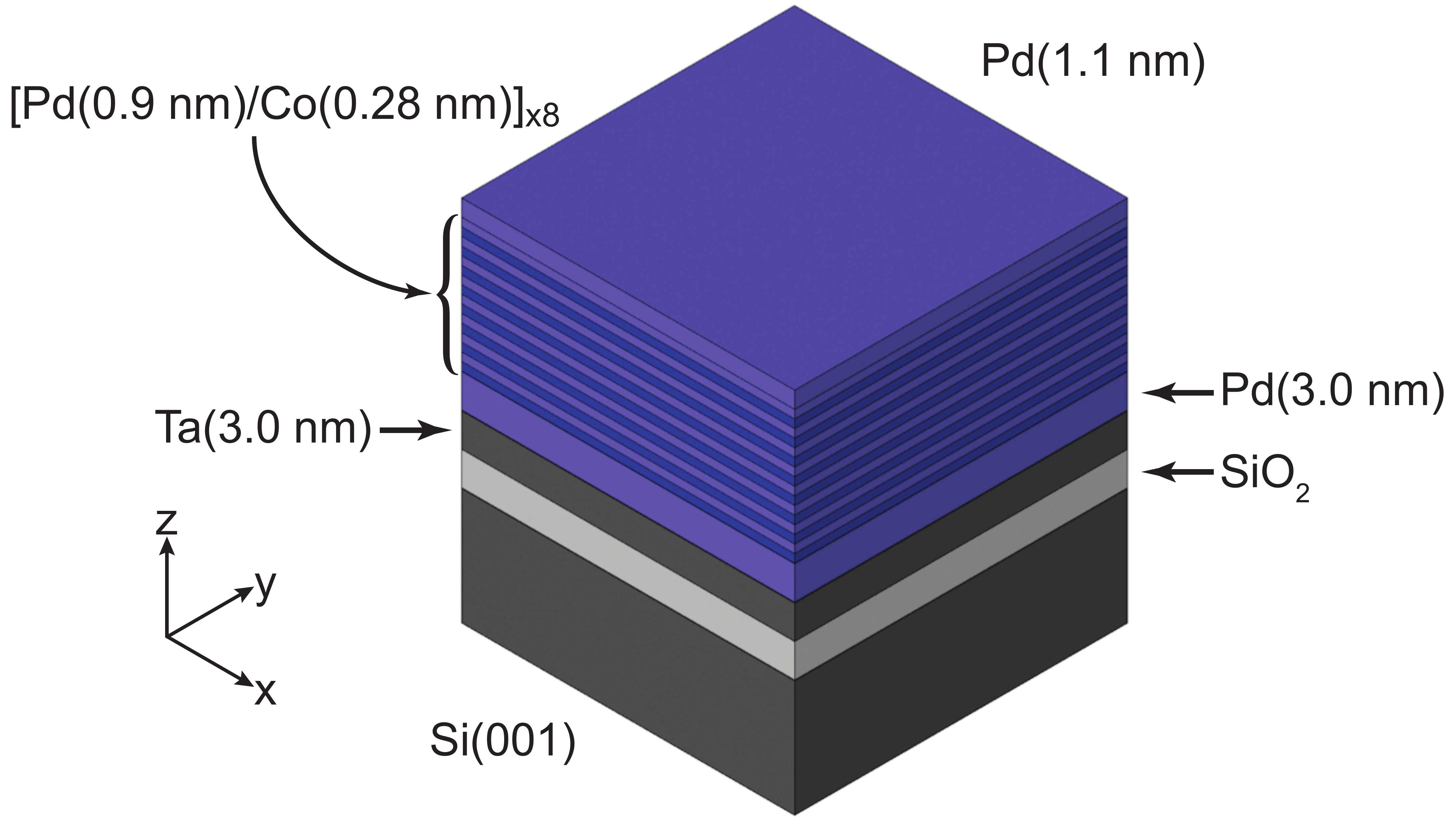}
 \caption{\label{Layeredfilm} Schematic of the different layers in the magnetic film used to fabricate the nanostructures (not to scale). The film structure is composed of Pd(1.1~nm) $+$ 8 $\times$ [Pd(0.9~nm)/Co(0.28~nm)] $+$ Pd(3.0~nm) $+$ Ta(3.0~nm) + SiO$_2$ $+$ Si(001).}
\end{figure}

A promising alternative that may overcome some of these limitations involves the use of magnetic lattice potentials~\cite{Ghanbari06,Gerritsma06,Xing07,Gerritsma07,Boyd07,Singh08,Whitlock09,Abdelrahman10,Llorente10,Leung11,Jose14}, in which patterned perpendicularly magnetized planar films create an array of magnetic microtraps able to trap ultracold atoms. In principle, such magnetic lattices provide robust potentials for manipulating atoms, combined with a high degree of design freedom, allowing arbitrary trap geometries and lattice spacings not restricted to certain fractions of the optical wavelengths. Other promising characteristics are the state selectivity of the traps (only atoms in weak-field seeking states remain trapped allowing manipulation using radiofrequency fields) and the possibility to incorporate other on-chip manipulation and detection devices.

The realization of arbitrary lattice patterns with sub-micron period would allow tunneling between the traps, ultimately enabling the study of exotic condensed matter
phenomena in nontrivial geometries, such as triangular, hexagonal, kagome, and superlattices, including honeycomb lattices. However, despite these potential benefits of magnetic lattices, no magnetic lattices with sub-micron period have been implemented to date.
 \begin{figure*}[Hbt!]%
\centering
\begin{subfigure}[b]{0.96\columnwidth}
\includegraphics[width=\columnwidth]{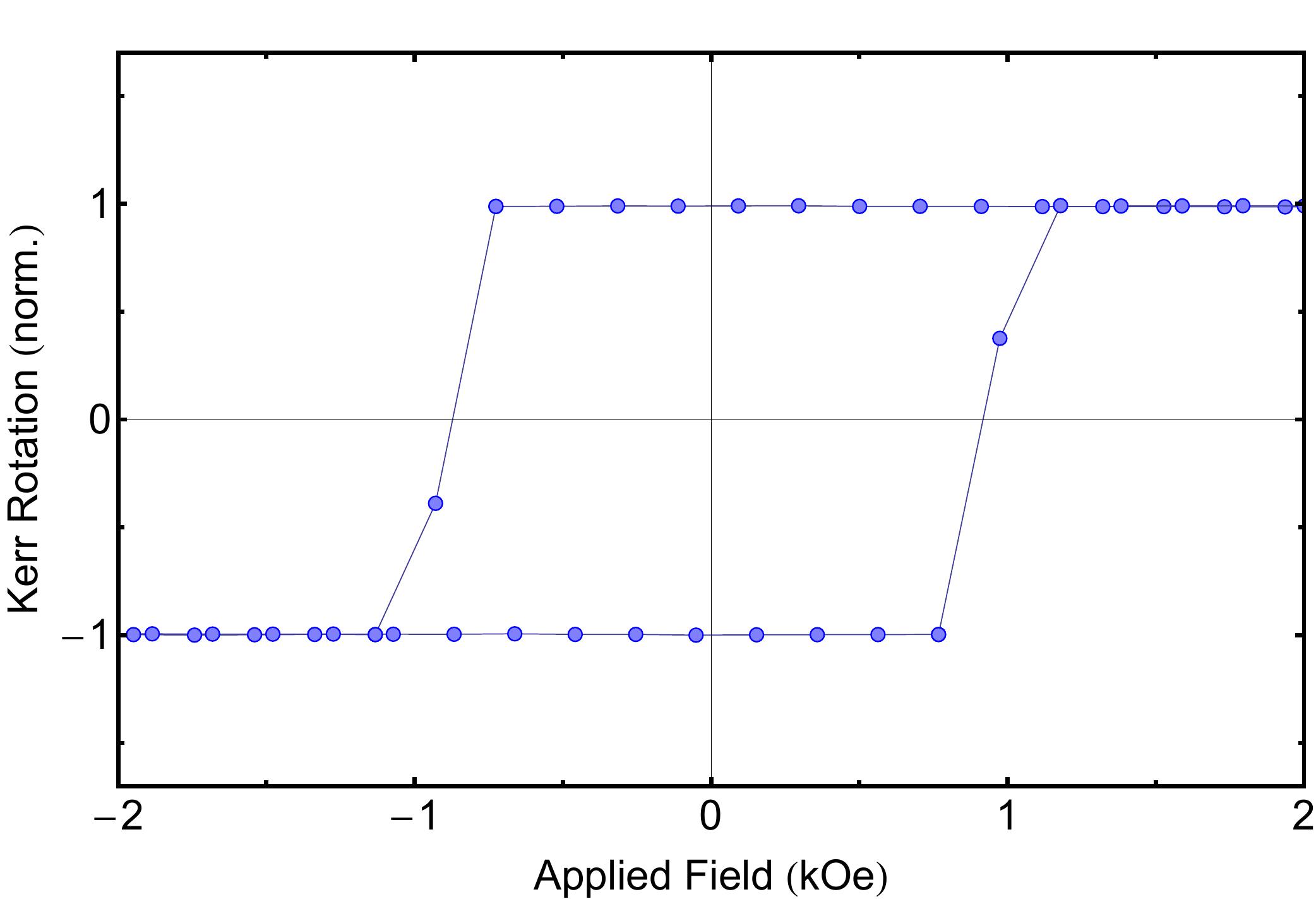}%
\caption{}%
\label{MOKE}%
\end{subfigure}\qquad%
\begin{subfigure}[b]{.80\columnwidth}
\includegraphics[width=\columnwidth]{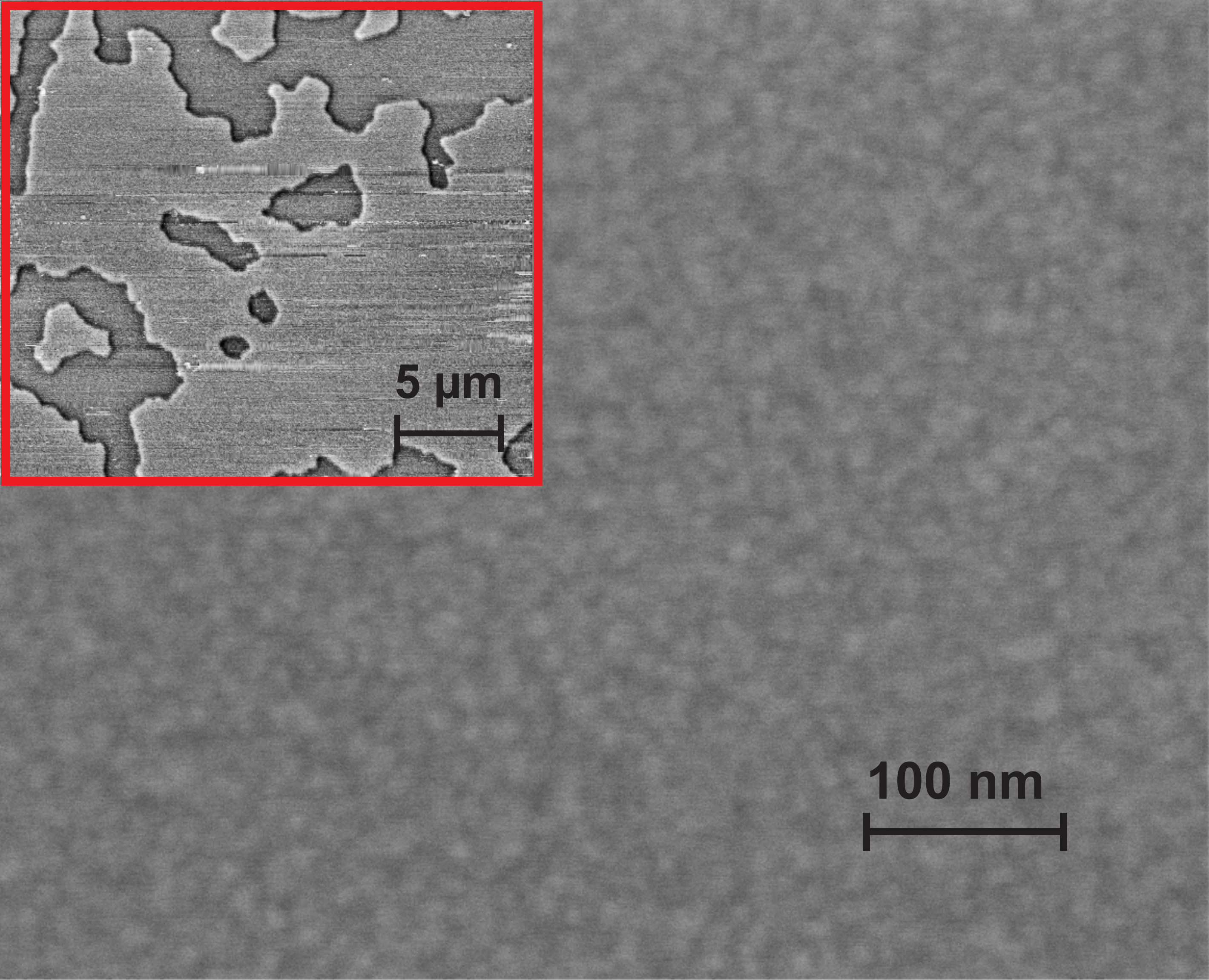}%
\caption{}%
\label{GrainSize}%
\end{subfigure}%
\caption{(a) Hysteresis loop of the Co/Pd multilayer film measured by Kerr rotation. (b) Scanning electron microscope (SEM) image of the surface of the Co/Pd film. From this image one can monitor the homogeneity of the material and the grain size. The measured average grain size is around 10~nm. Inset: Magnetic force microscope (MFM) image shows a Co/Pd multilayer in the demagnetized state revealing large magnetic domains with magnetization pointing up or down perpendicular to the film plane due to the strong PMA of this material.}
\end{figure*}

This paper describes the design, fabrication and characterization of sub-micron period one-dimensional (1D) and two-dimensional (2D) lattices of magnetic Ioffe-Pritchard microtraps for ultracold neutral atoms. The microtraps are produced near the surface of a patterned Co/Pd film with perpendicular magnetization. The magnetic lattices described here are a 1D lattice and two 2D lattices with square and triangular geometries with a period of $a=0.7$~$\mu$m. Scaling down the lattice period will increase the energy scales in the Hubbard model~\cite{Greiner02} used to describe quantum gases in optical lattices. The recoil energy, $E_{R}=(\pi \hbar)^{2}/2ma^2$ where $m$ is the atom mass, is used as a natural unit for these energy scales. Due to this dependence on the lattice period $a $ the tunneling rate $J$ and the on-site interaction $U $ can be scaled with the period of the lattice~\cite{Bloch08}, becoming large for sub-micron period lattices. In our case with a period of 0.7~$\mu$m and with a barrier height $V_{0}\thicksim 12 E_{R}$ or $6 E_{R}$, the critical tunneling rate is estimated to be $J_{C}\thicksim17$~Hz or 76~Hz~\cite{Bakr09}.

\section{Co/Pd magnetic films}

Recently, magnetic thin films, including multilayered systems, have been studied largely because of their interesting magnetic properties and possible applications for practical devices. In particular, studies of Co/Pd and Co/Pt~\cite{Carcia85,Broeder87,Draaisma87,Draaisma88,Carcia88,Sato88,Broedcr89,Ochiai89,Hashimoto90} systems have been performed extensively because they exhibit perpendicular magnetic anisotropy (PMA), small grain size and large Kerr rotation at short wavelengths, which makes them promising candidates for ultra-high density magnetic recording media~\cite{Grobis11} and also for atom optics research~\cite{Eriksson04}. We have chosen Co/Pd multilayer films for fabrication of our  magnetic sub-micron structures to trap ultracold atoms because of its high degree of homogeneity, high saturation magnetization and small grain size.

A strong PMA is necessary so that all the magnetic domains are aligned to give smooth and well defined magnetic potentials to trap the atoms on a sub-micron scale. At the same time, a high PMA leads to large coercivities necessary to withstand the external magnetic fields that are applied during the magnetic trapping of the atoms. In these multilayer films the large PMA is related to the reduced symmetry at the interface between the magnetic Co and the non-magnetic Pd layers. The magnetic film  consists of a set of bi-layers of Co/Pd, deposited on a 330~$\mu$m-thick silicon substrate, as shown in figure~\ref{Layeredfilm}. This bi-layered film is set on a layer of 3~nm-thick Pd to provide a good (111) texture to start the deposition of the magnetic layers, leading to an improvement of the crystallographic orientation of the layers and to an improvement of the PMA. The number of bilayers and the thickness of the layers were chosen to have high saturation magnetization and a large PMA; in this case we chose 8 bi-layers of alternating Co (0.28~nm) and Pd (0.9~nm). The Co/Pd multilayers were dc magnetron sputter-deposited at room temperature. The Ar pressure was adjusted to $3.5\times10^{-3}$~mbar for all depositions, while the base pressure of the deposition chamber was $1.0\times10^{-8}$~mbar. The measured coercivity for this film is $H_c=1.0$~kOe (see figure~\ref{MOKE}), and the saturation magnetization for 8 bi-layers of Co/Pd is $4 \pi M_s=5.9$~kG. One can increase the saturation magnetization by increasing the number of bi-layers but the PMA becomes less pronounced~\cite{Lin91} or may even vanish when using many layers. Finally, a layer of 1.1~nm of Pd is deposited on top of the stack to provide protection against oxidation (figure~\ref{Layeredfilm}).

The Co/Pd films can have a grain size down to about 6~nm~\cite{Roy11} and can produce high resolution magnetic structures necessary to prepare high quality 1D and 2D magnetic lattices with periods down to about 0.7~$\mu$m with smooth potentials. The grain size of the film was measured by scanning electron microscopy (SEM) (see figure~\ref{GrainSize}); the size of individual grains of Co/Pd within our samples is typically 10~nm. This small grain size currently sets the limit for the dimension of the smallest features we can etch.

\section{Patterning}

Sub-micron magnetic structures can be produced by different microfabrication techniques. These methods range from magneto-optical recording using a focused laser beam~\cite{Lau99,Eriksson04,Jaakkola05}, recording on a hard-disk head~\cite{Eriksson04,Boyd07}, use of grooved substrates plus uniform film coating \cite{Sidorov02,Wang05,Singh09,Jose14}, femtosecond laser ablation~\cite{Wolff09} and optical or e-beam lithography (EBL) followed by reactive ion etching (RIE)~\cite{Xing07B,Gerritsma07,Whitlock09}. We chose EBL and RIE since they provide high resolution and high versatility, arbitrary magnetization patterns are possible, and they can be used to produce the required sub-micron scales. 

\begin{figure}[htb]
 \centering
 \includegraphics[width=0.48\textwidth]{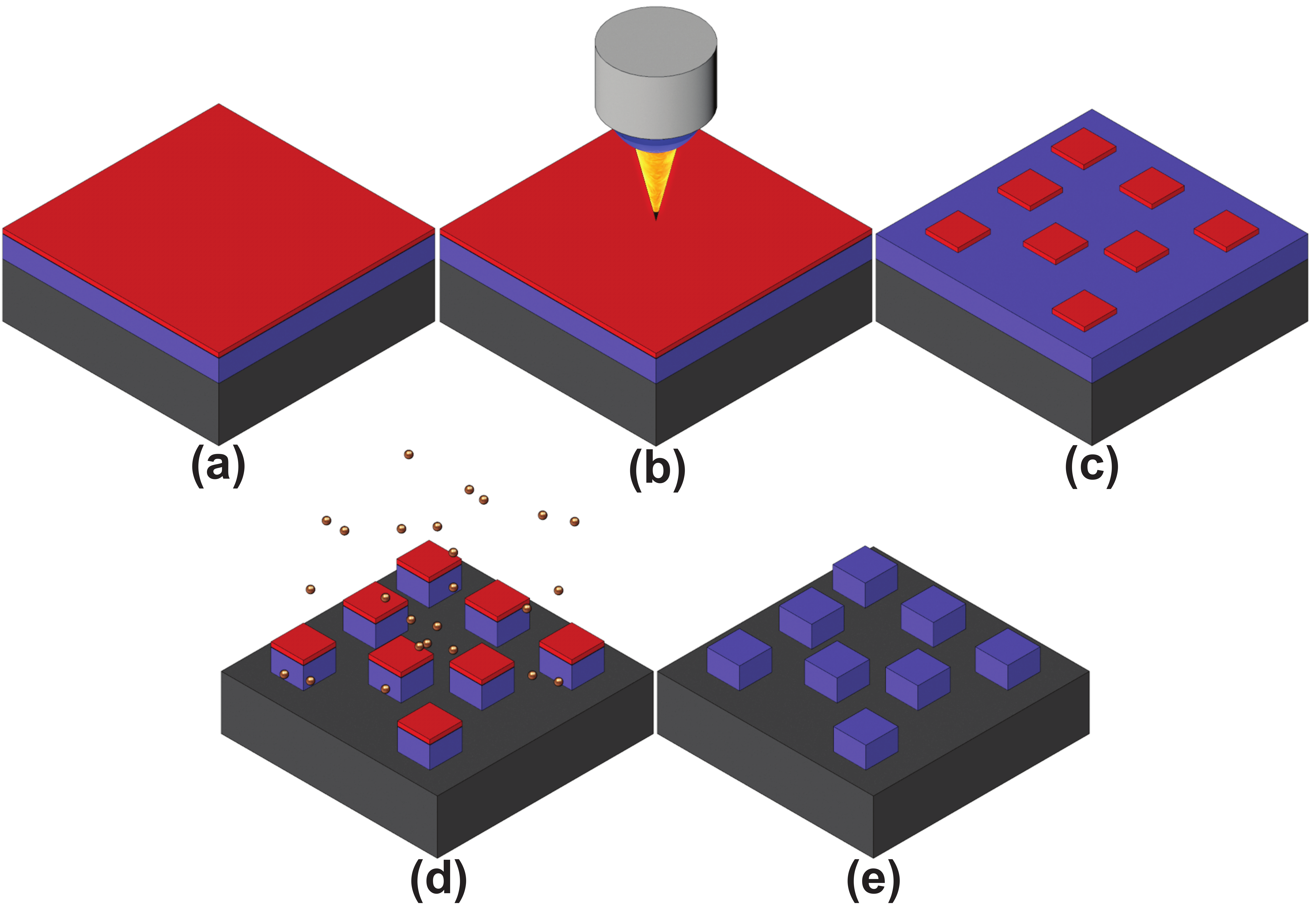}
 \caption{\label{ebeamlithography} Sequence of the process used to pattern the magnetic film. The colour label is: red resist, blue magnetic film, grey substrate. The process starts with (a) spin coating of PMMA positive resist, (b) EBL exposure, (c) resist development, (d) plasma etching by ion bombardment, and finally (e) removal of the remaining resist.}
\end{figure}

EBL was used to write the desired patterns onto the Co/Pd film. EBL is a maskless direct write method in which a beam of electrons is focused to expose required patterns on an electron-sensitive resist used as a mask for subsequent etching. The main advantage of EBL is that it can fabricate customized patterns with nanometer resolution; our EBL has 10~nm resolution sufficient for our patterns. The lithographic process was performed in the nano-fabrication facility at Swinburne University of Technology.
The lithography procedure is illustrated in figure~\ref{ebeamlithography}. First, a 300~nm-thick film of PMMA, a positive electron-sensitive resist is spin coated onto the Co/Pd film. Then a focused electron beam is driven by a high speed patterning generator on the mask with the desired design. After exposure of the resist layer to the electron beam this is developed and the exposed resist is removed. After that the magnetic film is plasma-etched by RIE with argon. This is a dry etching technique with good properties for high resolution patterning, no resist adhesion problems and high anisotropic etch profile~\cite{Manos89}. In the final stage the remaining resist is removed by wet etching, using acetone. After the patterning process a metallic reflecting layer of 50~nm thickness is deposited over the surface; this layer is necessary for creating a mirror-MOT. Finally, a layer of silica (25~nm) is deposited to prevent rubidium atoms from sticking to the metallic surface.

\section{Results for 1D structures}

\subsection{Trapping parameters}

A 1D lattice potential of magnetic microtraps can be created by a periodic array of  long, parallel rectangular magnets with out-of-plane magnetization in the presence of suitable bias magnetic fields~\cite{Ghanbari06,Gerritsma06,Boyd07,Singh08}. By superimposing on this potential homogeneous in-plane bias fields in the parallel and perpendicular direction, one can control most of the important parameters related to the atom trapping, such as the distance at which the trap minima are generated from the surface, the trap frequencies and the bias field necessary to have non-zero potential minima at the bottom of the traps to avoid spin-flips. For an infinite 1D lattice periodic in the $y $ direction contained in the $xy $ plane and for distances from the surface large compared with $a/2\pi$ (where $a $ is the lattice period), the trapping magnetic fields can be expressed as~\cite{Singh09}\

\begin{figure*}[htb!]%
\centering
\begin{subfigure}{0.96\columnwidth}
\includegraphics[width=\columnwidth]{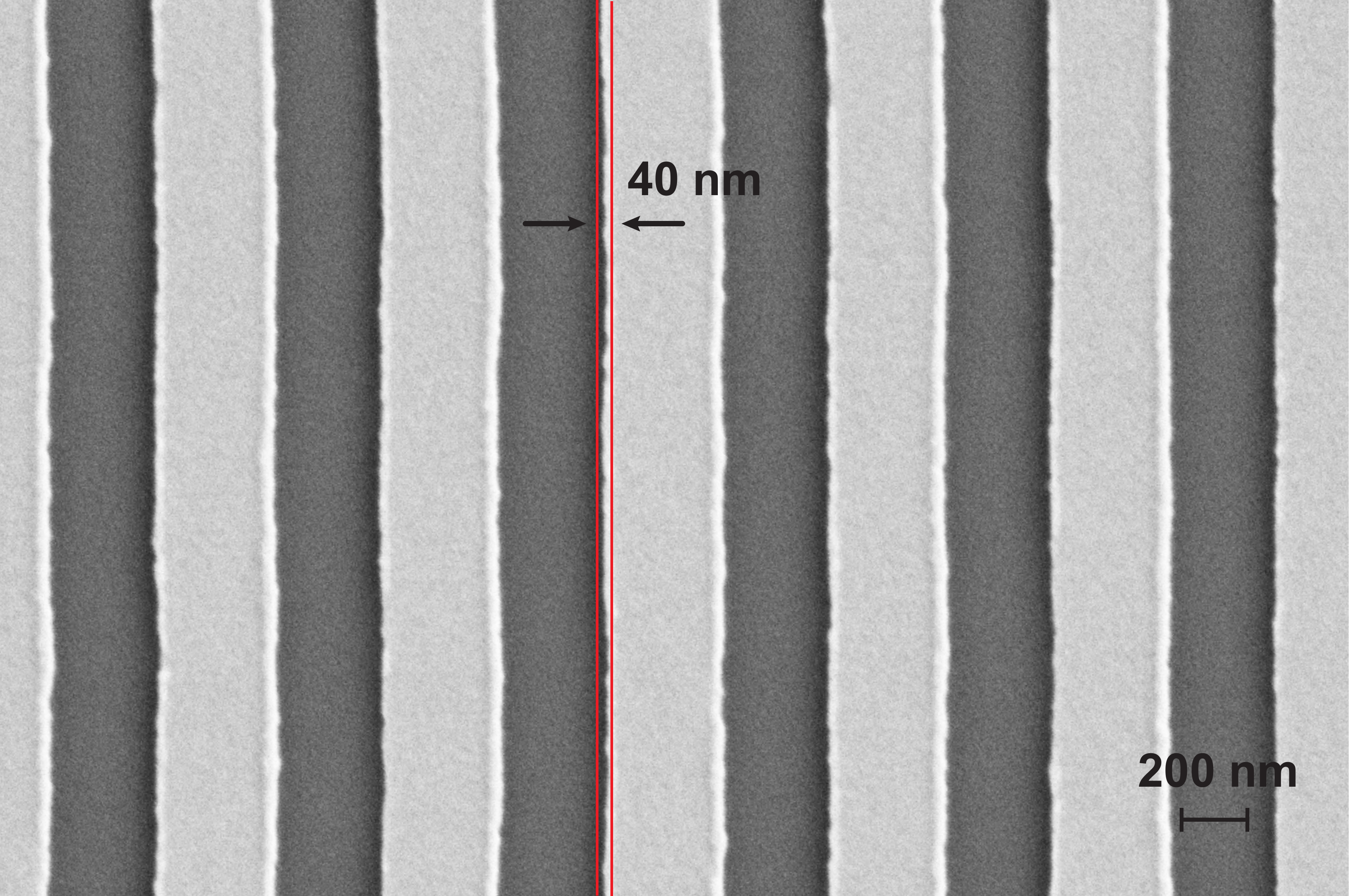}%
\caption{}%
\label{1DSEM1}%
\end{subfigure}\quad%
\begin{subfigure}{.96\columnwidth}
\includegraphics[width=\columnwidth]{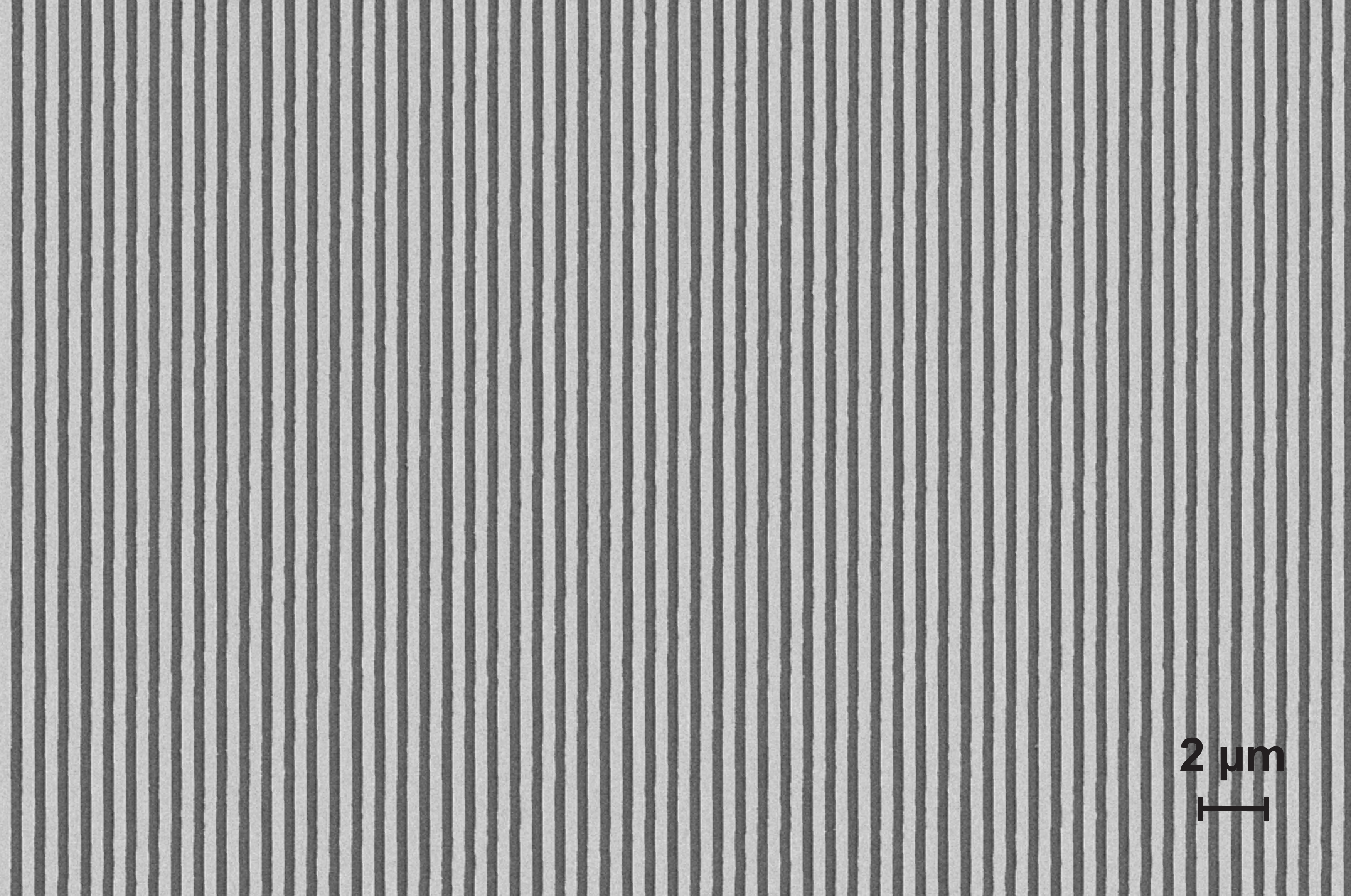}%
\caption{}%
\label{1DSEM2}%
\end{subfigure}%
\caption{(a) Small-scale SEM image of the 1D structure. The dark regions correspond to the etched part and the bright regions the magnetic film. The image shows a pattern of parallel trenches with a period close to 0.7~$\mu$m. From the image we observe an edge roughness with an amplitude of 40 nm. (b) Large-scale SEM image. A Fourier transform of the SEM image reveals that the peaks are mainly multiples of a base frequency of 688 nm corresponding to the period of the structure. The FWHM of the base frequency peak is 7~nm and the heights of other peaks are less than 1~\%.}
\end{figure*}

\begin{equation}\label{1Dfield}
[B_{x},B_{y},B_{z}]=[B_{bx},B_{0}sin(ky)e^{-kz}+B_{by},B_{0}cos(ky)e^{-kz}], 
\end{equation}
where $k=2\pi/a$, $B_{0}=4M_{z}(e^{kt}-1)$ (Gaussian units), $t $ is the magnetic film thickness, $ M_{z}$ is the magnetization and $B_{bx}$, $B_{by}$ are the bias magnetic fields in the 
$x $ and $y $ directions. From this equation one can calculate the relevant magnitudes of the trapping potential as a function of the bias fields. The distance to the surface of the minima is
\begin{equation}
z_{min}=\frac{a}{2\pi}ln\left( \frac{B_{0}}{\left|B_{by}\right|}\right)
\end{equation}
The barrier heights are
\begin{equation}
\Delta B_{y}=(B_{bx}^2+4B_{by}^2)^{1/2}-\left|B_{bx}\right|\\
\textnormal{and}\\
\Delta B_{z}=(B_{bx}^2+B_{by}^2)^{1/2}-\left|B_{bx}\right|
\end{equation}

The Ioffe field at the bottom of the minima is $B_{IP}$=$\left|B_{bx}\right|$. The trapping potential is produced by the magnetic dipole interaction, which if the atom adiabatically follows the trapping potential can be expressed as $U(x,y,z)=m_{F}g_{F}\mu_{B} B$. Only atoms in weak-field seeking states can be trapped in a local minimum of the magnetic field, and in order to minimize three-body losses that limit the lifetime of the atoms in the tight microtraps, for $^{87}$Rb we choose the weak-field seeking state $\left|F=1,m_{F}=-1\right\rangle$. 
Then the radial trap frequency associated with a harmonic potential in elongated traps can be expressed as
\begin{equation}
\omega_{y}\approx\omega_{z}=\frac{2\pi}{a}\left(\frac{m_{F}g_{F}\mu_{B}}{m\left|B_{bx}\right|}\right)^{1/2}B_{by}
\end{equation}

From this equation we can estimate some characteristics of the trapping potential associated with our magnetic film patterned as a 1D array with a period 0.7~$\mu$m. The magnetic film has $4 \pi M_{z}=5.9$~kG and the total magnetic thickness is $t=2.24$~nm. One can create a set of microtraps with $z_{min}=280$~nm, where the required bias field is $B_{by}=3$~G. To avoid Majorana spin-flips the bottom of the trap needs to be raised from zero; if the bottom of the trap is set to $B_{IP}=1$~G, the barrier heights are $\Delta B_{y}=5.1$~G and  $\Delta B_{z}=2.2$~G. For $^{87}$Rb atoms in the $ \left|F=1,m=-1 \right\rangle$ state, these barrier heights translate into trap depths of 171~$\mu$K and 73~$\mu$K, respectively, which are sufficient to trap atoms pre-cooled to 10-15~$\mu$K. The trap frequencies in this case are $\omega_{y}/ 2\pi \approx \omega_{z}/ 2\pi \approx 100$~kHz.

Atom surface interactions can limit the lifetime of the atom clouds at distances from the surface of the order of a hundred nanometers. Previous experimental studies show short lifetimes at distances of about 1~$\mu$m~\cite{Harber03,Lin04}. The main reasons for this limit are the van der Waals forces and Johnson noise. In the first case the attractive van der Waals force between the atoms and the surface can shift and alter the magnetic potential so that it is no longer trapping, as happens with the gravity potential if the trap is not tight enough. To compare both forces one can define a critical trap frequency~\cite{Leung11} at which the trapping potential begins to fold due to the van der Waals forces. If the interacting surface is the final reflecting layer and using the previous settings of the potential we can calculate the critical trap frequency, $\omega_{crit}/2\pi=44$~kHz. This frequency value is below the trapping frequency, from which we conclude that the stiffness of the magnetic potential is sufficiently large to compensate the van der Waals force.

Johnson noise comes about because a conducting surface can have random thermal currents flowing in-plane; these currents are a source of magnetic field noise which can induce spin flips and hence loss of atoms from a magnetic trap~\cite{Henkel99,Henkel01,Harber03,Lin04,Rekdal04,Scheel05}. For thin conducting films such that the film thickness $t << z_{min} << \delta $, where $\delta$=$\sqrt{2\rho/ (\mu_{0} \omega_{L}})$ is the skin depth (typically 50-100~$\mu$m) at the spin-flip transition frequency $\omega_L$, the thermal spin-flip transition rate can be written~\cite{Lin04,Scheel05}
\begin{equation}\label{Jlifetime}
\Gamma_{Fm}\approx C_{Fm}^{2} C_0 t /z_{min}^2
\end{equation}
where $C_{Fm}^2 =\left| \left\langle F,m+1\left| S_+\right|F,m\right\rangle \right|^2$, $S_+$ is the electron-spin raising operator, $C_0=\left[ 3\mu_0 g_s \mu_B/\left(8 \hbar\right)\right]^2\left[k_B T /\left(6 \pi \rho \right)\right]=$\\
68~$\mu $m~s$^{-1}(T/300$~K$)(\rho_{Au}/\rho)$~\cite{Lin04}, $\rho$ is the resistivity of the conducting film ($2.22 \times 10^{-8}~\Omega$m for gold), $\mu_0$ is the vacuum permeability, $\omega_L = g_F \mu_{B} B_{IP}/\hbar$ is the Larmor frequency, $g_F$ is the Lande g-factor, $g_S \approx 2$ is the electron spin g-factor and $\mu_B$ is the Bohr magneton. Thus for $t << z_{min} << \delta $ the spin-flip lifetime $\tau=\Gamma_{Fm}^{-1}$ scales approximately as $z_{min}^2 \rho/t$, which is independent of the $\omega_L$ and hence the trap bottom $B_{IP}$. For $^{87}$Rb atoms trapped in the $ \left|F=1,m=-1 \right\rangle$ state, atom loss is assumed to occur via the $ \left|F=1,m=-1 \right\rangle \rightarrow \left|F=1,m=0 \right\rangle$ transition, for which $C_{1,-1}^2 = 1/8$. For a gold reflecting film and $t=0.05$~$\mu$m, $z_{min}=0.28$~$\mu$m, $T=300$~K, we obtain a thermal spin-flip lifetime $\tau_{Au} \approx 180$~ms which is long compared with the estimated tunneling times of 60~ms and 13~ms for a 0.7~$\mu$m-period lattice with barrier heights $V_0 \sim 12 E_{R}$~(20~mG) and $6 E_{R}$~(10~mG), respectively. The spin-flip lifetime could be increased if required by using a reflecting film with higher resistivity, such as palladium ($\rho_{Pd}= 1.05 \times 10^{-7}$~$\Omega$m), for which the spin-flip lifetime becomes $\tau_{Pd} \approx 870$~ms.
 
\begin{figure*}[htb!]%
\centering
\begin{subfigure}[b]{0.80\columnwidth}
\includegraphics[width=\columnwidth]{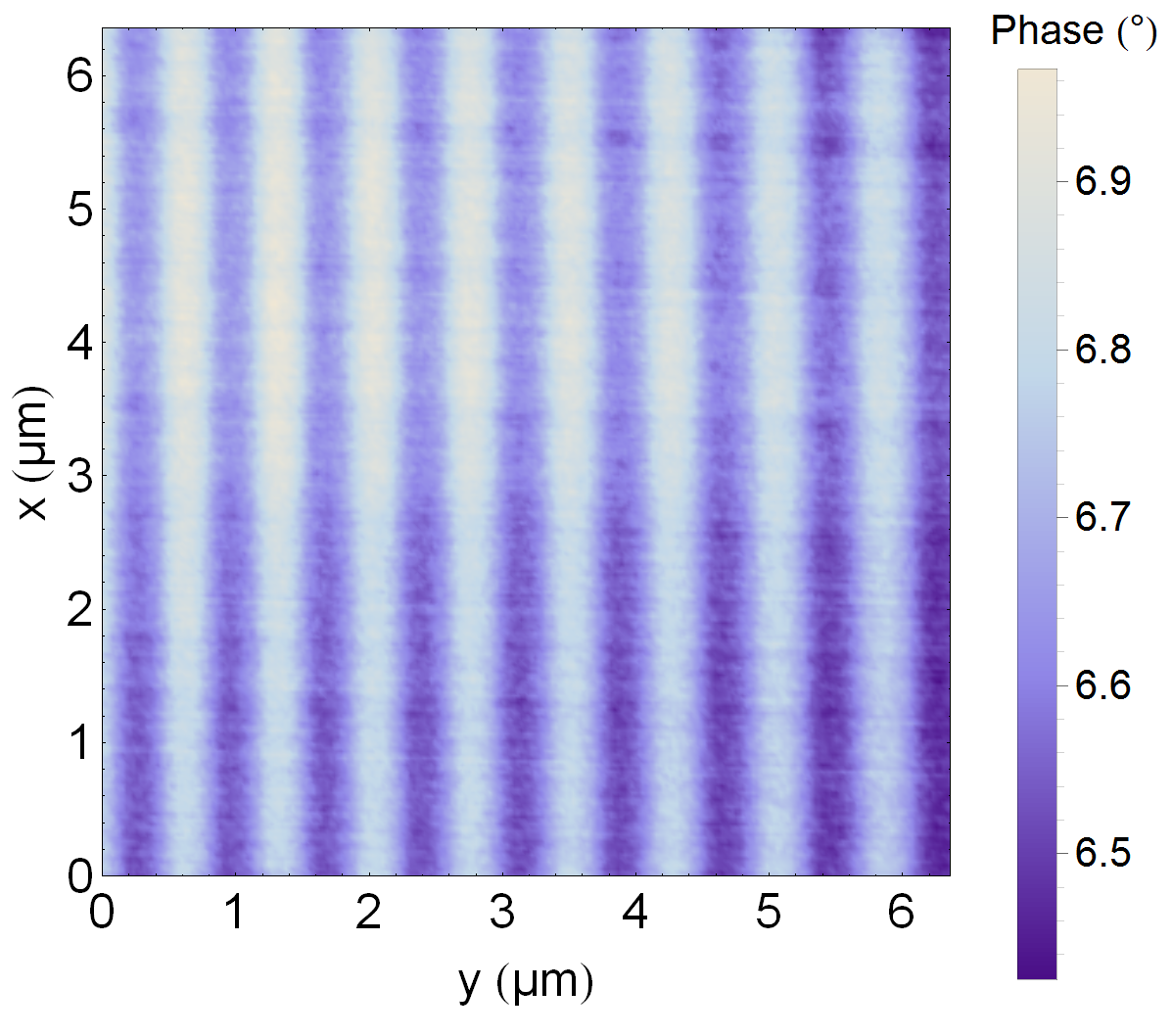}%
\caption{}%
\label{PhaseMap}%
\end{subfigure}\quad%
\begin{subfigure}[b]{1.1\columnwidth}
\includegraphics[width=\columnwidth]{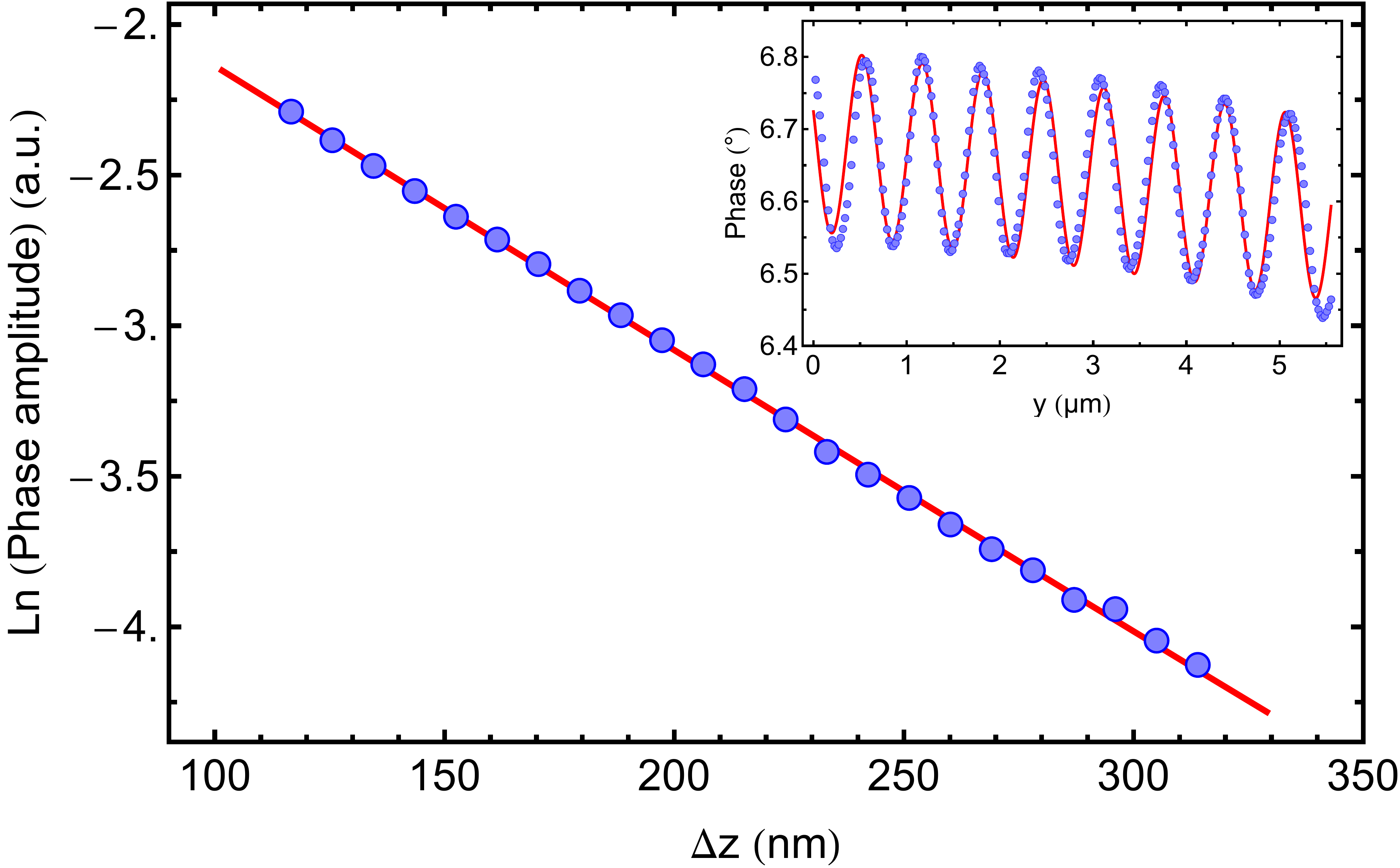}%
\caption{}%
\label{Decay}%
\end{subfigure}%
\caption{(a) MFM measurement of the 1D magnetic microstructure with a distance between the probe tip and the surface of 50~nm. The MFM image maps the magnetic field above the surface of the 1D sample, which shows oscillations given by the period of the structure. (b) Plot of the natural logarithm of the amplitude of the MFM signal versus tip-surface distance $\Delta z$. The red line is a fitted decay curve. Inset (b): Plot of the profile of the MFM signal in the $y$ direction at a tip surface distance of 50~nm. The red line is the fitted curve to the data points from the oscillating signal.}
\end{figure*}
 
\subsection{Fabrication results}

To obtain these 1D structures a square region of dimension 0.8$\times $0.8~mm${^2}$ of the magnetic film  was etched following the method described in figure~\ref{ebeamlithography}. The resulting film was examined with an SEM (figures~\ref{1DSEM1} and~\ref{1DSEM2}) to check the quality of the surface and the period of the grooved structures. A magnetic force microscope (MFM) scan over the magnetized sample was carried out to map the magnetic field over the structures~(figure~\ref{PhaseMap}) This scan is made in the so-called dynamic MFM mode (AC) to increase the signal to noise. However, the resulting magnitude does not give us directly the magnetic field over the surface. The AC MFM mode provides a measure of the difference of the phase of the oscillating cantilever-sample system. The tip oscillates at its resonant frequency with a small amplitude in the vertical $z$ direction. To lowest order the magnetic force causes a phase shift and a shift in the resonant frequency~\cite{Grutter92,Manalis95}
\begin{eqnarray}
\Delta\phi\approx \frac{Q}{k}\frac{\partial F_z}{\partial z} \propto \frac{\partial^2 B_z}{\partial z^2}, \quad\quad
\Delta f\approx -\frac{f_n}{2k}\frac{\partial F_z}{\partial z} \propto \frac{\partial^2 B_z}{\partial z^2}
\end{eqnarray}
where $Q$ is the cantilever quality factor, $f_n$ is the natural resonant frequency of the cantilever tip and $k$ is the spring constant. Thus the MFM signal is primarily sensitive to the second spatial derivative of the $z$ component of the magnetic field. If we use the magnetic field from equation~\ref{1Dfield}, the MFM signal is just proportional to the magnetic field, an oscillating signal in the $y $ direction with the period of the structures and an amplitude decay length of $k^{-1}=a/2\pi$. A  way to check the quality of the magnetic lattice across the sample is to determine the dependence of the amplitude of this oscillating MFM signal on the distance of the MFM tip from the etched magnetic film. From the oscillating profile of the MFM (figure~\ref{Decay} inset, taken at a tip-surface distance of 50~nm), we measure a period of $a_{osc}=651$~$\pm$~3~nm, and from the fitted decay length (figure~\ref{Decay}) we obtain $a_{decay}=662$~$\pm$~11~nm, where the uncertainties come from the residuals of the fits. These values are close to the result from the SEM analysis (figure~\ref{1DSEM2}) confirming the quality of the periodicity of the structure. Some of the difference between the two measurements comes from the calibration of the spatial dimensions in the MFM apparatus or from high harmonic terms that we have neglected in this analysis~\cite{Hughes97}.

\begin{figure*}[htb!]%
\centering
\begin{subfigure}{0.96\columnwidth}
\includegraphics[width=\columnwidth]{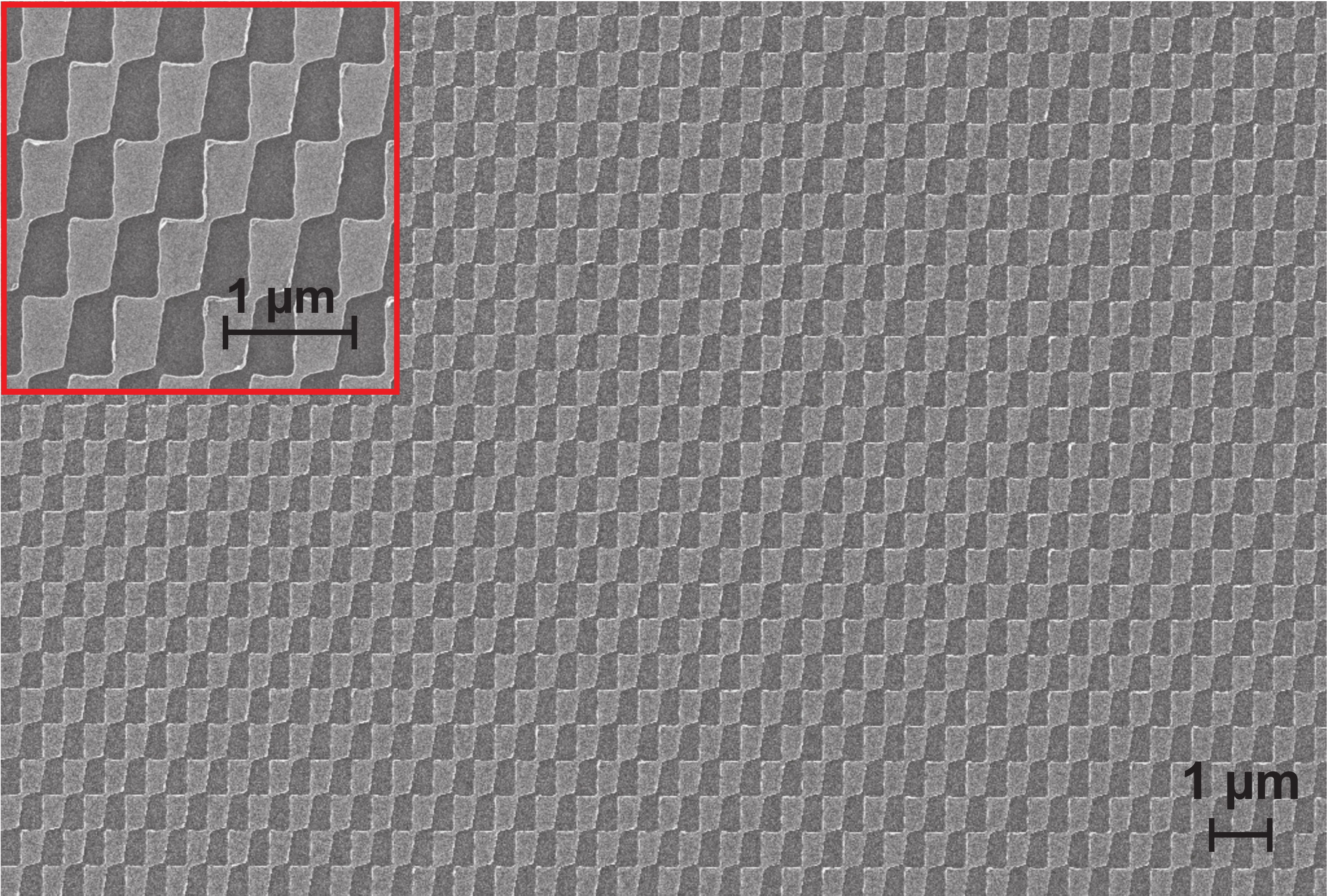}%
\caption{}%
\label{SEMTriangular}%
\end{subfigure}\quad%
\begin{subfigure}{.96\columnwidth}
\includegraphics[width=\columnwidth]{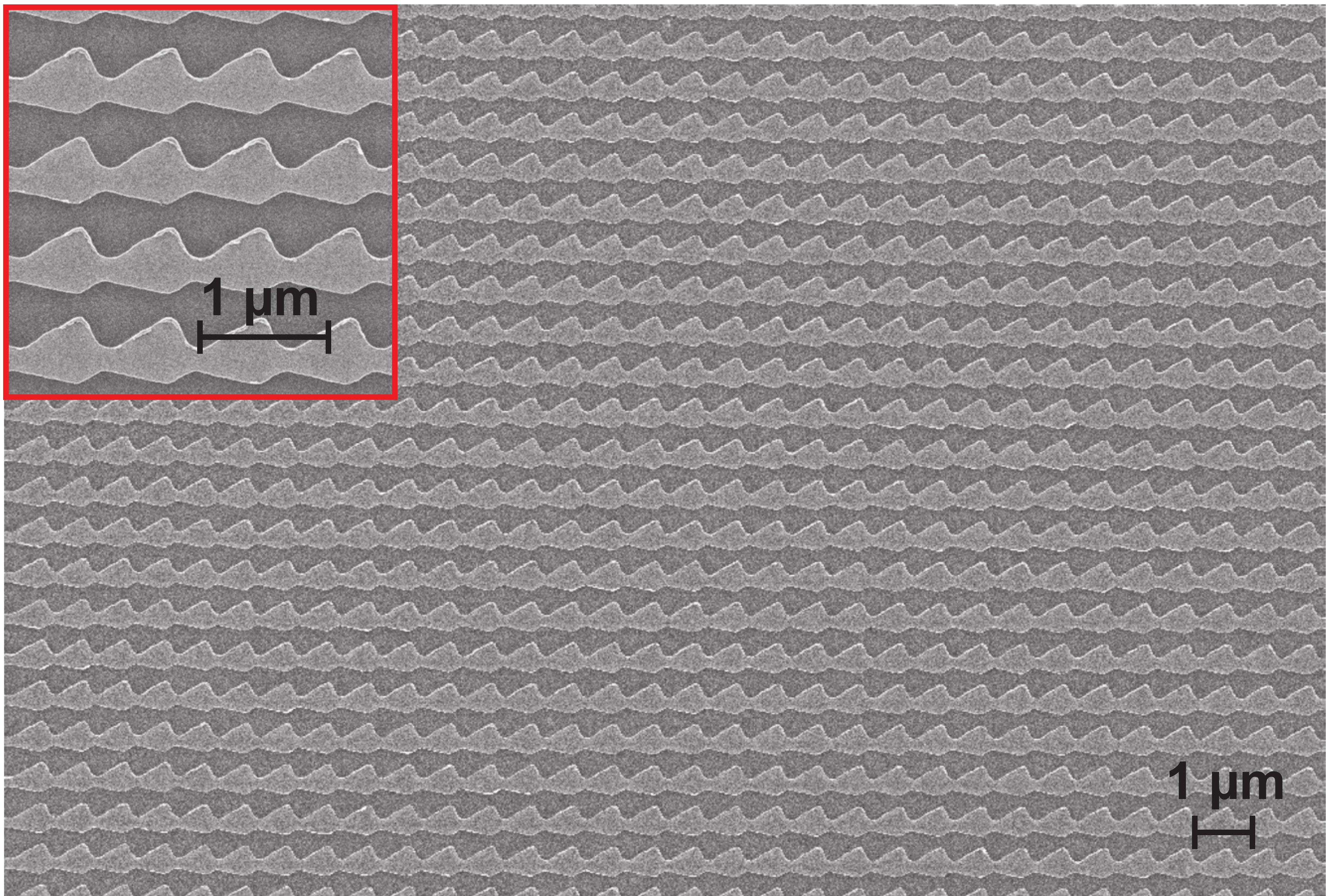}%
\caption{}%
\label{SEMSquare}%
\end{subfigure}%
\caption{SEM images of the patterns of 2D geometries after the etching process. The dark grey regions correspond to the etched and the light grey magnetic regions are the non-etched. These patterns generate a magnetic lattice with (a) triangular translational symmetry, and (b) square translational symmetry. Both structures exhibit good homogeneity (from the main SEM picture) and the desired single element block which is duplicated across the film (from the inset at the top left of each image)}
\end{figure*}

\section{Results for 2D structures}
 
The fabrication of magnetic structures to create two-dimensional complex magnetic lattices with sub-micron period for trapping ultracold atoms is the main goal of the present work. Until now 2D lattices with square, triangular, kagome or honeycomb geometries have been created with optical lattices~\cite{Tarruell12,Uehlinger13,Stamper13,Luhman14}, as well as square and triangular lattice geometries for permanent magnetic films~\cite{Whitlock09,Leung14} but with periods of 10~$\mu$m or greater. We have fabricated two magnetic planar structures able to create magnetic potentials with square and triangular symmetries and sub-micron period.
 
In principle, a planar structure etched in a magnetic film can create arbitrary potentials above the surface~\cite{Schmied10}. To design the magnetic structures related to the desired potentials we have employed an optimization algorithm proposed by Schmied et al.~\cite{Schmied10}. The code automatically generates a planar binary distribution of magnetization patterns with magnetization out of plane, in order to produce the desired lattice symmetries with specified trap parameters. This method makes possible the design of the desired geometries, which would be extremely difficult to obtain using manual methods. The optimization algorithm itself exports a binary image which encodes the magnetic versus non-magnetic regions within one unit cell, so that the generated pattern has pixels with either zero or maximal magnetization. Once a magnetic lattice unit cell has been designed it is necessary to export it in a format for lithographic patterning; then the lithographic software will tile the unit cell to produce the entire lattice. The fabrication process at this point is straightforward due to the binary nature of the pattern generated, by etching away the pixels with zero magnetization.
 
Our simulations indicate that for a square magnetic lattice produced by a patterned Co/Pd film with $4 \pi M_z=5.9$~kG and $t=2.24$~nm, bias fields of $B_{bx}=1.7$~G, $B_{by}=-0.8$~G create lattice traps at $z_{min} = 350$~nm from the film surface, with a Ioffe field of $B_{IP}=1.1$~G. The trap barrier height is 1.42~G in both the $x $ and $y $ directions and 0.78~G in the $z$ direction. For $^{87}$Rb atoms in the $ \left|F=1,m=-1 \right\rangle$ state, these barrier heights correspond to trap depths of 47~$\mu$K and 26~$\mu$K, respectively. Each trap is approximately cylindrically symmetric with the long axis along the $[1,1,0] $ direction. The calculated trap frequencies are $\omega_{z}/2\pi \approx \omega_{\perp}/2\pi \approx 120$~kHz and $\omega_{\parallel}/2 \pi \approx 37$~kHz.

For the triangular lattice, bias fields of $B_{bx}=0.1$~G, $B_{by}=-1.0$~G create traps at $z_{min} = 350$~nm from the film surface, with a Ioffe field of $B_{IP}=0.36$~G. The trap barrier height is 1.3~G along all three triangular lattice directions and 0.62~G in the $z$ direction. These barrier heights correspond to trap depths of 43~$\mu$K and 21~$\mu$K, respectively. Larger barrier heights can be obtained by using thicker Co/Pd magnetic film. The trap frequencies are $\omega_{x}/2\pi \approx 36$~kHz and $\omega_{y}/2\pi \approx \omega_{z}/2\pi \approx 145$~kHz. For $z_{min}=0.35$~$\mu$m from a gold or palladium layer with $t=0.05$~$\mu$m, we obtain from equation~\ref{Jlifetime} thermal spin-flip lifetimes $\tau_{Au}=290$~ms and $\tau_{Pd}=1360$~ms.
 
The resulting etched structures for these geometries are shown in figure~\ref{SEMTriangular} and~\ref{SEMSquare}. The measured etch depth is around 10~nm, so that we can be certain that all of the magnetic film is removed in the desired non-magnetic zones for a good reproduction of the binary pattern. From the SEM images we observe good agreement with the designed pattern, and from a far field of view the unit cell is well reproduced for large areas~$>100 \times 100$~$\mu$m$^2$. Some stitching errors appear in the pattern due to the limited area of patterning of the EBL. An effective working area of 100~$\mu$m$^2$ is satisfactory for our purpose, since this corresponds to more than 10$^4$ lattice sites in a 0.7~$\mu$m-period lattice.

\begin{figure*}%
\centering
\begin{subfigure}{.48\columnwidth}
\includegraphics[width=\columnwidth]{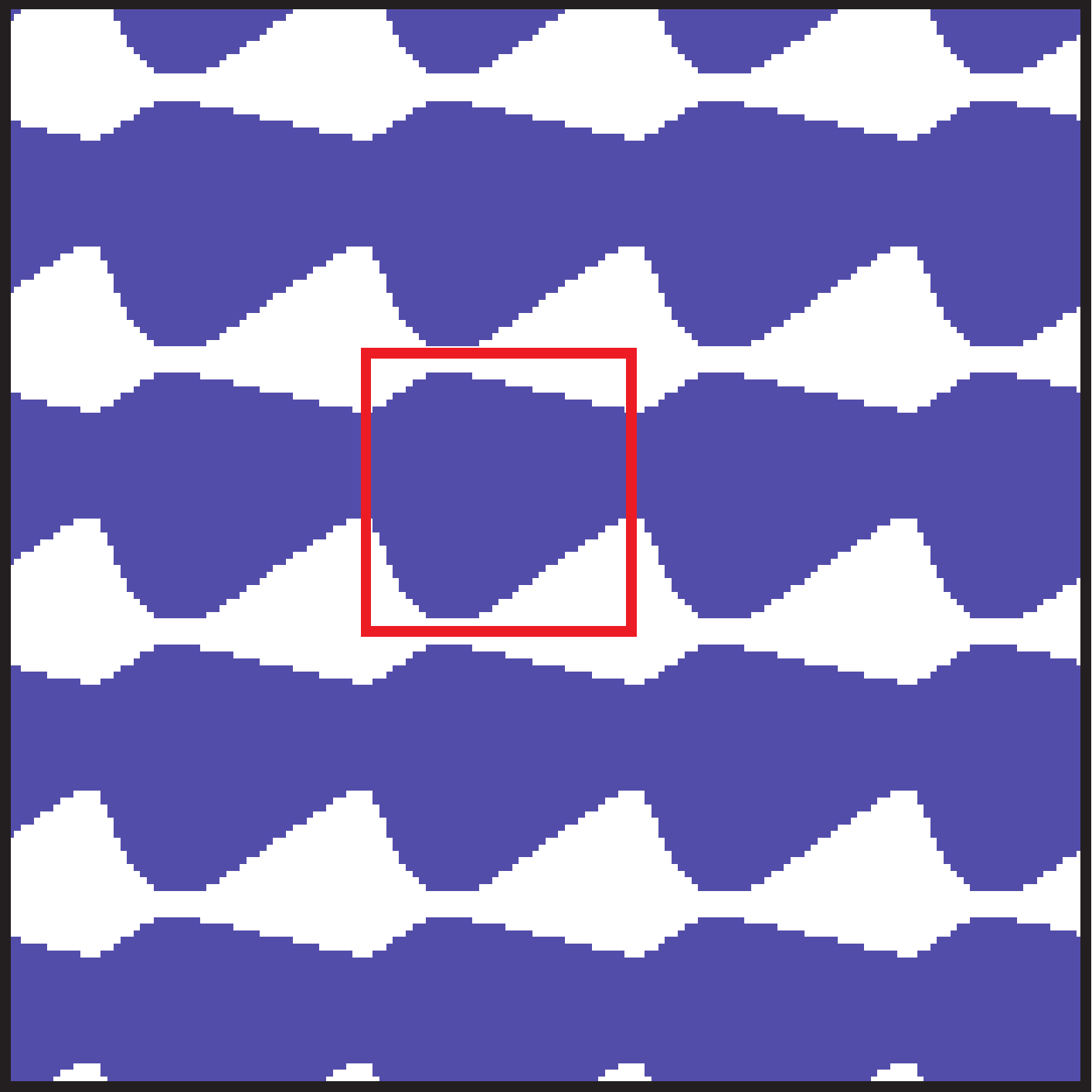}%
\caption{}%
\label{MFMTestA}%
\end{subfigure}\quad%
\begin{subfigure}{.48\columnwidth}
\includegraphics[width=\columnwidth]{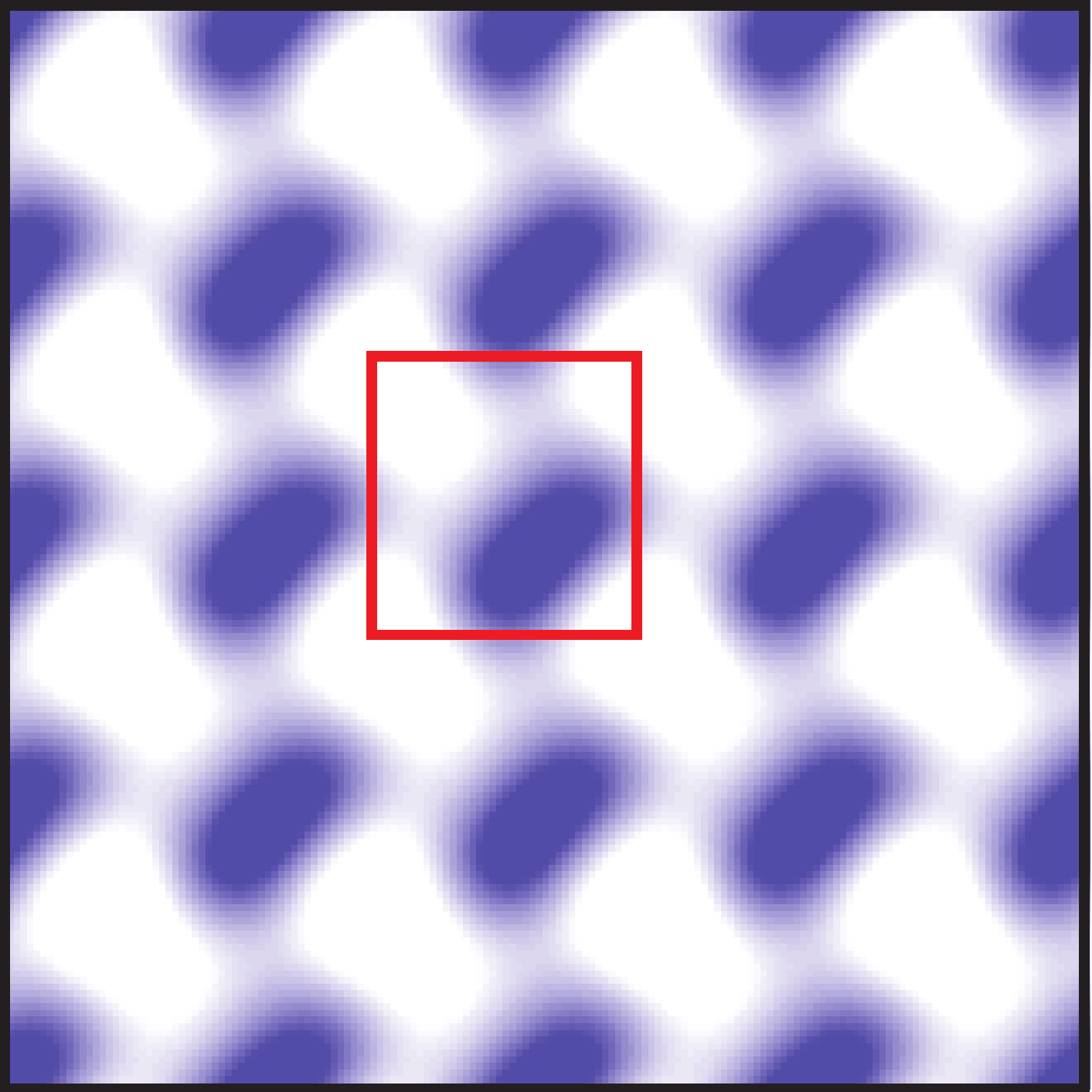}%
\caption{}%
\label{MFMTestB}%
\end{subfigure}\quad%
\begin{subfigure}{.48\columnwidth}
\includegraphics[width=\columnwidth]{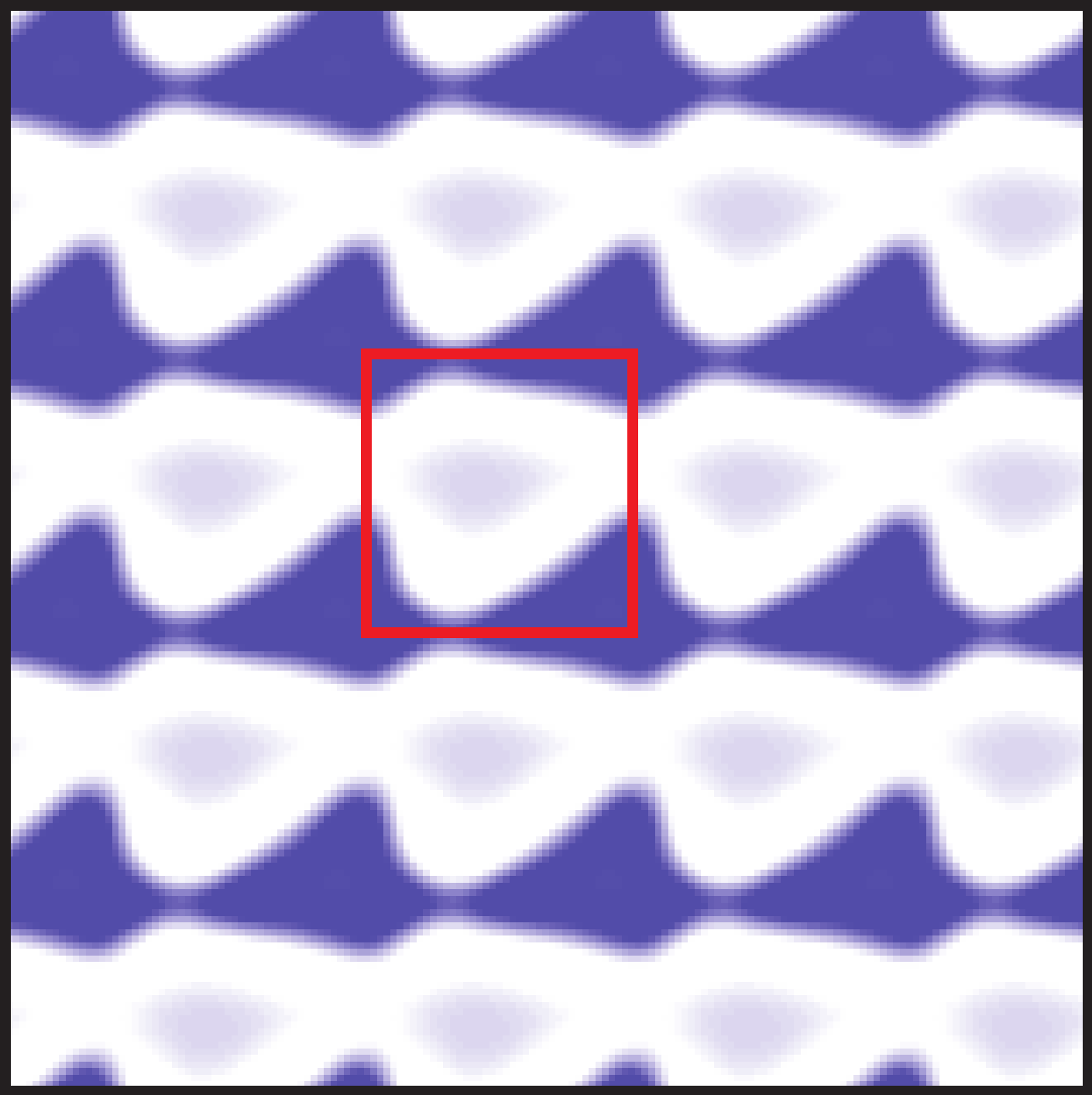}%
\caption{}%
\label{MFMTestC}%
\end{subfigure}\quad%
\begin{subfigure}{.48\columnwidth}
\includegraphics[width=\columnwidth]{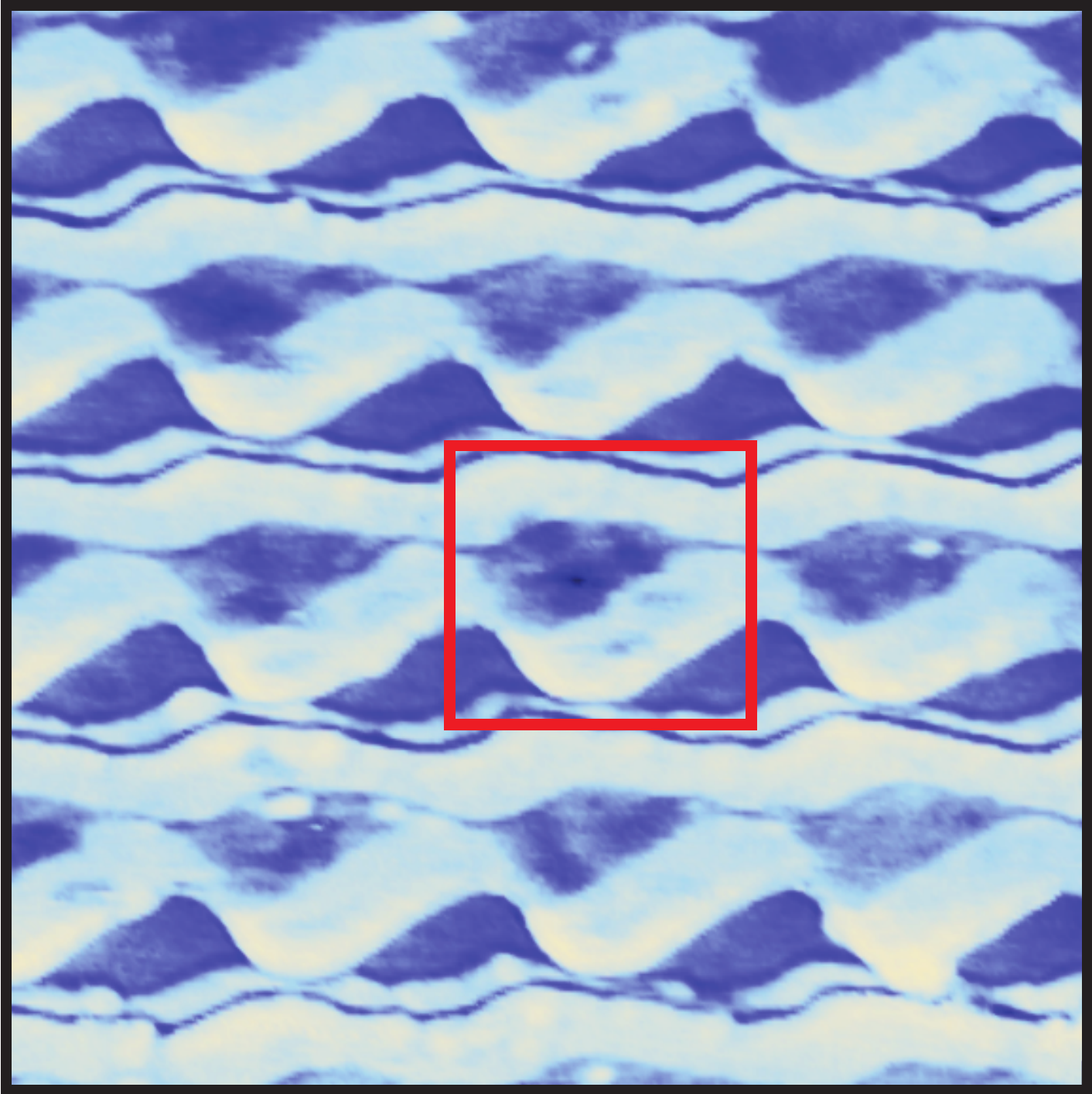}%
\caption{}%
\label{MFMTestD}%
\end{subfigure}%
\caption{(a) Pattern generated by the Schmied et al code~\cite{Schmied10}. The white region corresponds to the magnetic film. The red square is delimiting the unit cell. (b) Calculated magnetic potential of a square magnetic lattice of period $a=$~0.7~$\mu$m at the  chosen height of $a/2$ produced by the planar magnetic structure. (c) Calculated second spatial derivative of $B_z$ at a constant height calculated from the designed structures. (d) MFM  signal from the fabricated structure, in qualitative agreement with the second spatial derivative. For this calculation (in (a), (b) and (c)) an extra erosion in the film during the patterning was taken into account.}
\label{MFMTest}
\end{figure*}

We have performed calculations of the magnetic field produced by the optimized square lattice, and in particular for the second derivative of the field for comparison with the MFM data \cite{Hughes97}. Figure~\ref{MFMTest} shows the surface topology and calculated field for the optimized square lattice. The calculated second derivative is in qualitative agreement with the MFM
measurements of the lithographically patterned Co/Pd multilayer. In the MFM data there are several sharper (nearly) horizontal lines within the magnetic islands that are not reproduced by the simulations. This feature could be explained by the formation of magnetic domains near the boundaries of the structure, or by redeposition of material at one edge during etching. Presently, we cannot rule out artifacts introduced by the MFM when the surface topology changes due to residual photoresist on the surface.
 
\section{Summary}
 
We have fabricated and characterized magnetic structures to create a lattice of magnetic microtraps for ultracold atoms with a period of 0.7~$\mu$m. We fabricated one 1D lattice and two 2D lattices with square and triangular translational geometries. To create the structures we have patterned a Co/Pd magnetic multilayered thin film with electron-beam lithography and reactive ion etching. We have shown that Co/Pd multilayered films have a small grain size and high remanent magnetization and coercivity making them an ideal material to create sub-micron magnetic structures. The scanning electron microscope  images displayed the required patterning quality and magnetic force microscope measurements confirm that the structures have the right magnetic potentials to trap ultracold atoms in the diverse lattice geometries.  New developments in 3D electron beam and ion beam technologies free from stitching errors may allow us to combine resolution of tens-of-nm with areas having sub-cm cross-sections.
These small periods give access to high energy scales in the Hubbard model and to tight traps with high frequencies for rubidium atoms. This makes them a promising tool for studies of one-dimensional quantum gases, quantum simulation and quantum information processing.
The method demonstrated in this paper offers a great degree of freedom to design and create different magnetic potentials to trap ultracold atoms close to surfaces. Geometries that are difficult to generate with other methods, such as triangular, hexagonal, kagome, and superlattices, may provide a better understanding of the underlying physics of novel materials such as graphene~\cite{Zhu07,Tarruell12,Uehlinger13}, a remarkable material that has great technological potential and is currently the subject of intense interest.

\phantomsection
\section*{Acknowledgments} 

\addcontentsline{toc}{section}{Acknowledgments} 

We are indebted to James Wang, Stefan Tibus, Brenton Hall, Russell Anderson and Smitha Jose for fruitful discussions. We acknowledge funding from an Australian Research Council (ARC) Discovery Project grant (Grant No. DP130101160).
\phantomsection\

\bibliographystyle{unsrt}
\bibliography{SubmicronAtomchip}

\begin{thebibliography}{10}

\bibitem{Jaksch98PRL}
Jaksch D. Bruder C. Cirac J.~I. Gardiner~C. W. and Zoller P.
\newblock {\em Phys. Rev. Lett}, 81:3108, 1998.

\bibitem{Morsch06}
Morsch O. and Oberthaler M.
\newblock {\em Rev. Mod. Phys}, 78:179, 2006.

\bibitem{Bloch08}
Bloch I.~Dalibard J. and Zwerger W.
\newblock {\em Rev. Mod. Phys}, 80:885, 2008.

\bibitem{Paredes04}
Paredes~B. et~al.
\newblock {\em Nature}, 429:277, 2004.

\bibitem{Kinoshita06}
Wenger~T Kinoshita~T and Weiss~D S.
\newblock {\em Nature}, 440:900, 2006.

\bibitem{Cataliotti01}
Cataliotti F.~S. Burger S. Fort C. Maddaloni P. Minardi F. Trombettoni
  A.~Smerzi A. and Inguscio M.
\newblock {\em Nature}, 293:843, 2001.

\bibitem{Albiez05}
Albiez M. Gati R. F\"olling J. Hunsmann S.~Cristiani M. and Oberthaler M.
\newblock {\em Phys. Rev. Lett.}, 95:010402, 2005.

\bibitem{Greiner02}
Greiner M. Mandel O. Esslinger T.~Hansch T. and Bloch I.
\newblock {\em Nature}, 415:39, 2002.

\bibitem{Calarco00}
Calarco T. Hinds E.~A. Jaksch D. Schmiedmayer J. Cirac~J. I. and Zoller P.
\newblock {\em Phys. Rev. A.}, 61:022304, 2000.

\bibitem{Monroe07}
Monroe C.
\newblock {\em Nature}, 416:469, 2007.

\bibitem{Lewenstein07}
Lewenstein M. Sanpera A. Ahufinger V. Damski B. Sen~De A. and Sen U.
\newblock {\em Adv. Phys.}, 56:243, 2007.

\bibitem{Ghanbari06}
Ghanbari S. Kieu T.~D.~Sidorov A. and Hannaford P.
\newblock {\em J. Phys. B}, 39:847, 2006.

\bibitem{Gerritsma06}
Gerritsma R. and Spreeuw R.~J. C.
\newblock {\em Phys. Rev. A.}, 74:043405, 2006.

\bibitem{Xing07}
Xing T. Barb I. Gerritsma R. Spreeuw R.~J.~C. Luigjes H. Xiao Q.~F.~Retif C.
  and Goedkoop~J. B.
\newblock {\em J. Magn. Magn. Mater.}, 313:192--197, 2007.

\bibitem{Gerritsma07}
Gerritsma R. Whitlock S. Fernholz T. Schlatter H. Luigjes J.~A. Thiele J.~U.
  Goedkoop~J. B. and Spreeuw R.~J. C.
\newblock {\em Phys. Rev. A}, 76:033408, 2007.

\bibitem{Boyd07}
Medley P. Campbell G.~K. Mun J. Ketterle~W. Boyd M. Streed E.~W. and
  Pritchard~D. E.
\newblock {\em Phys. Rev. A}, 76:043624, 2007.

\bibitem{Singh08}
Singh M. Volk M. Akulshin A. Sidorov A.~McLean R. and Hannaford P.
\newblock {\em J. Phys. B}, 41:065301, 2008.

\bibitem{Whitlock09}
Whitlock S. Gerritsma R.~Fernholz T. and Spreeuw R.~J. C.
\newblock {\em New J. Phys.}, 11:023021, 2009.

\bibitem{Abdelrahman10}
Abdelrahman A. Vasiliev M.~Alameh K. and Hannaford P.
\newblock {\em Phys. Rev. A}, 82:012320, 2010.

\bibitem{Llorente10}
Llorente Garcia I. Darquie B. Curtiss E.~A. Sinclair C.~D. J. and Hinds~E. A.
\newblock {\em New J. Phys.}, 12:093017, 2010.

\bibitem{Leung11}
Leung V.~Y.~F. Tauschinsky~A. van Druten N.~J. and Spreeuw R.~J. C.
\newblock {\em Quantum Inf. Process.}, 10:955, 2011.

\bibitem{Jose14}
Jose S. Surendran P. Wang Y. Herrera I. Krzemien L. Whitlock S. McLean
  R.~Sidorov A. and Hannaford P.
\newblock {\em Phys. Rev. A (R)}, 89:051602, 2014.

\bibitem{Bakr09}
Bark W.~S. et~al.
\newblock {\em Nature}, 462:74, 2009.

\bibitem{Carcia85}
Carcia P.~F. Meinhaldt~A. D. and Suna A.
\newblock {\em Appl. Phys. Lett.}, 47:174, 1985.

\bibitem{Broeder87}
den Broeder F.~J.~A. Donkersloot H.~C. Draaisma H.~J.~G. and de~Jonge W.~J.~M.
\newblock {\em J. App. Phys.}, 61:4317, 1987.

\bibitem{Draaisma87}
Draaisma H.~J.~G. de~Jonge W.~J.~M. and den Broeder F.~J.~A.
\newblock {\em J. Magn. Magn. Mat.}, 66:351, 1987.

\bibitem{Draaisma88}
Draaisma H.~J.~G. den Broeder F.~J.~A. and de~Jonge W.~J.~M.
\newblock {\em J. Appl. Phys.}, 63:3479, 1988.

\bibitem{Carcia88}
Carcia~P. F.
\newblock {\em J. Appl. Phys.}, 63:51166, 1988.

\bibitem{Sato88}
Sato N.
\newblock {\em J. App. Phys.}, 64:6424, 1988.

\bibitem{Broedcr89}
den Broedcr F.~J.~A. Kuiper D. Donkersloot H.~C. and Hoving W.
\newblock {\em Appl. Phys. A}, 49:507, 1989.

\bibitem{Ochiai89}
Ochiai Y.~Hashimoto S. and Aso K.
\newblock {\em IEEE Trans. Magn.}, 25:3755, 1989.

\bibitem{Hashimoto90}
Hashimoto S.~Matsuda H. and Ochiai Y.
\newblock {\em App. Phys. Lett.}, 56:1069, 1990.

\bibitem{Grobis11}
Grobis M.~K. Hellwig O. Hauet T.~Dobisz E. and Albrecht~T. R.
\newblock {\em IEEE Trans. Magn.}, 47:6, 2011.

\bibitem{Eriksson04}
Eriksson S. Ramirez-Martinez F. Curtis E.~A. Sauer B.~E. Nutter P.~W. Hill~E.
  W. and Hinds~E. A.
\newblock {\em Appl. Phys. B}, 79:811--816, 2004.

\bibitem{Lin91}
Lin C-J. Gorman G.~L. Lee C.~H. Farrow R.~F.~C. Marinero E.~E. Do H.~V.~Notarys
  H. and Chien~C. J.
\newblock {\em J. Magn. Magn. Mat.}, 93:194--204, 1991.

\bibitem{Roy11}
Roy A.~G. Laughlin D.~E. Klemmer T.~J. Howard K.~Khizroev S. and Litvinov D.
\newblock {\em J. Appl. Phys.}, 89:11, 2011.

\bibitem{Lau99}
Lau D.~C. McLean R.~J. Sidorov A.~I. Gough D.~S. Koperski J. Rowlands W.~J.
  Sexton B.~A. Opat~G. I. and Hannaford P.
\newblock {\em J. Opt. B: Quantum Semiclass. Opt.}, 1:371, 1999.

\bibitem{Jaakkola05}
Jaakkola A. Shevchenko A. Lindfors K. Hautakorpi M. Ilyashenko E. Johansen~T.
  H. and Kaivola M.
\newblock {\em Eur. Phys. J. D}, 35:81, 2005.

\bibitem{Sidorov02}
Sidorov A. McLean R.~J. Scharnberg F. Gough D.~S. Davis T.~J. Sexton B.~J.
  Opat~G. I. and Hannaford P.
\newblock {\em Acta Phys. Polon.}, B33:2137, 2002.

\bibitem{Wang05}
Wang J.~Y. Whitlock S. Scharnberg F. Gough D.~S. Sidorov A.~I. McLean~R. J. and
  Hannaford P.
\newblock {\em J. Phys. D: Appl. Phys.}, 38:4015, 2005.

\bibitem{Singh09}
Singh M. McLean R.~Sidorov A. and Hannaford P.
\newblock {\em Phys. Rev. A}, 79:053407, 2009.

\bibitem{Wolff09}
Wolff C.~H. Whitlock S. Lowe R.~M. Sidorov~A. I. and Hall~B. V.
\newblock {\em J. Phys. B: At. Mol. Opt. Phys.}, 42:085301, 2009.

\bibitem{Xing07B}
Xing Y.~T. Barb I. Gerritsma R. Spreeuw R.~J.~C. Luigjes H. Xiao Q.~F.~Rétif
  C. and Goedkoop~J. B.
\newblock {\em J. Magn. Magn. Mat.}, 313:192, 2007.

\bibitem{Manos89}
Manos D. and Flamm D.
\newblock {\em Plasma Etching an Introduction Academic Press Inc. New York},
  1989.

\bibitem{Harber03}
Harber et~al.
\newblock {\em J. Low Temp. Phys.}, 133:229, 2003.

\bibitem{Lin04}
Lin Y.~J. Teper I.~Chin C. and Vuletic V.
\newblock {\em Phys. Rev. Lett.}, 92:050404, 2004.

\bibitem{Henkel99}
Henkel C.~P\"otting S. and Wilkens M.
\newblock {\em Appl. Phys. B}, 69:379, 1999.

\bibitem{Henkel01}
Henkel C. and P\"otting S.
\newblock {\em Appl. Phys. B}, 72:73, 2001.

\bibitem{Rekdal04}
Rekdal P.~K. Scheel S. Knight~P. L. and Hinds~E. A.
\newblock {\em Phys. Rev. A.}, 70:013811, 2004.

\bibitem{Scheel05}
Rekdal P.~K. Knight P.~L. Scheel~. and Hinds~E. A.
\newblock {\em Phys. Rev. A.}, 72:042901, 2005.

\bibitem{Grutter92}
Grutter~P. et~al.
\newblock Scanning tunneling microscopy ii.
\newblock {\em edited by Wiensendanger R and Guntherodt. H J}, pages 151--207,
  1992.

\bibitem{Manalis95}
Manalis~S. et~al.
\newblock {\em Appl. Phys. Lett.}, 66:2585, 1995.

\bibitem{Hughes97}
Hughes I.~G. Barton P.~A. Roach~T. M. and Hinds~E. A.
\newblock {\em J. Phys. B: At. Mol. Opt. Phys.}, 30:2119, 1997.

\bibitem{Tarruell12}
Tarruell L. Greif D. Uehlinger T.~Jotzu G. and Esslinger T.
\newblock {\em Nature}, 483:302, 2012.

\bibitem{Uehlinger13}
Uehlinger T. Jotzu G. Messer M. Greif D. Hofstetter W.~Bissbort U. and
  Esslinger T.
\newblock {\em Phys. Rev. Lett.}, 111:185307, 2013.

\bibitem{Stamper13}
Parameswaran S.~A. Kimchi I. Turner A.~M. Stamper-Kurn~D. M. and Vishwanath A.
\newblock {\em Phys. Rev. Lett.}, 110:125301, 2013.

\bibitem{Luhman14}
L\"uhman D.~S. J\"urgensen O. Weinberg M. Simonet J. Saltan-Panahi P. and
  Sengstock K.
\newblock {\em Phys. Rev. A}, 90:013614, 2014.

\bibitem{Leung14}
Leung V.~Y.~H. et~al.
\newblock {\em Rev. Sci. Instrum.}, 85:053102, 2014.

\bibitem{Schmied10}
Schmied R. Leibfried D. Spreeuw R.~J. C. and Whitlock S.
\newblock {\em New J. Phys.}, 12:103029, 2010.

\bibitem{Zhu07}
Zhu S.~L.~Wang B. and Duan~L. M.
\newblock {\em Phys. Rev. Lett.}, 98:260402, 2007.

\end{thebibliography}


\end{document}